\documentclass[preprint]{raa}
\usepackage{amsmath}
\usepackage{afterpage}
\usepackage{placeins}
\usepackage{graphicx,times}             
\usepackage{natbib}
\usepackage{amssymb,amsmath}
\bibpunct{(}{)}{;}{a}{}{,}

\usepackage[pagebackref=true, colorlinks=true, allcolors=blue]{hyperref}

\usepackage[T1]{fontenc} 
\usepackage{mathptmx}
\usepackage{tabularx}
\usepackage{txfonts}
\usepackage{xspace}
\usepackage{gensymb} 
\usepackage{numprint}
\usepackage{xfrac}
\usepackage{orcidlink}

\npthousandsep{\,}

\def\hi{\textsc{Hi}\xspace}
\def\hiim{\textsc{Hi\xspace IM}\xspace}
\def\camb{\textsc{camb}\xspace}

\def\mpc{\rm Mpc}

\newcommand{\reffg}[1]{Figure~\ref{#1}}

\newcommand{\refeq}[1]{Equation~(\ref{#1})}
\newcommand{\refsc}[1]{Section~\ref{#1}}

\def\apjl{Astrophysical Journal Letters}
\def\aap{Astronomy and Astrophysics}

\def\apjs{The Astrophysical Journal Supplement Series}

\begin{document}

\title{Exploring \hi Galaxy Redshift Survey Strategies for the FAST Core Array Interferometry}

\volnopage{Vol.25 (2025) No.5, 000--000}      
\setcounter{page}{1}          

\author{Zhenglong Li \orcidlink{0009-0004-0677-3474} \inst{1} 
    \and 
    Diyang Liu \orcidlink{0009-0000-6895-9136} \inst{1} 
    \and 
    Chengliang Xu \orcidlink{0009-0008-6933-358X} \inst{2,3} 
    \and 
    Yichao Li$^\star$ \orcidlink{0000-0003-1962-2013} \inst{1} 
    \and 
    Xin Zhang \orcidlink{0000-0002-6029-1933} \inst{1,4,5}}

\institute{
Key Laboratory of Cosmology and Astrophysics (Liaoning) \& College of Sciences, Northeastern University,  Shenyang 110819, China; $^\star${\it liyichao@mail.neu.edu.cn}\\
    \and
    National Astronomical Observatories, Chinese Academy of Sciences, Beijing 100012, China;\\
    \and
    School of Astronomy and Space Science, University of Chinese Academy of Sciences, Beijing 100049, China;\\
    \and
    National Frontiers Science Center for Industrial Intelligence and Systems Optimization, Northeastern University, Shenyang 110819, China; \\
    \and
    Key Laboratory of Data Analytics and Optimization for Smart Industry (Ministry of Education), Northeastern University, Shenyang 110819, China.
    \vs\no
    {\small Received 2025 March 14; accepted 2025 April 9}}

\abstract{
We explore the feasibility of \hi galaxy redshift surveys with the Five-hundred-meter Aperture Spherical Telescope (FAST) and its proposed Core Array interferometry. Using semi-analytical simulations, we compare the performance of the FAST single-dish and Core Array modes in drift scan (DS) and on-the-fly (OTF) observations across different redshifts.  
Our results show that the FAST single-dish mode enables significant \hi detections at low redshifts ($z \lesssim 0.35$) but is limited at higher redshifts due to shot noise. The Core Array interferometry, with higher sensitivity and angular resolution, provides robust \hi galaxy detections up to $z \sim 1$, maintaining a sufficient number density for power spectrum measurements and BAO constraints. At low redshifts ($z \sim 0.01$–$0.08$), both configurations perform well, though cosmic variance dominates uncertainties. At higher redshifts ($z > 0.35$), the Core Array outperforms the single-dish mode, while increasing the survey area has little impact on single-dish observations due to shot noise limitations.  
The DS mode efficiently covers large sky areas but is constrained by Earth's rotation, whereas the OTF mode allows more flexible deep-field surveys at the cost of operational overhead. Our findings highlight the importance of optimizing survey strategies to maximize FAST’s potential for \hi cosmology. The Core Array is particularly well-suited for high-redshift \hi galaxy surveys, enabling precise constraints on large-scale structure and dark energy. 
\keywords{surveys -- cosmology: large-scale structure of Universe --- cosmology: observations
--- methods: observational}
}

\authorrunning{Z. Li, et. al. }            
\titlerunning{Exploring \hi Galaxy Redshift Survey Strategies for the FAST Core Array Interferometry}  

\maketitle

\section{Introduction} \label{sec:intro}

Hydrogen is the most abundant element in the universe. After the epoch of reionization (EoR), 
the neutral hydrogen (\hi) is widely distributed within galaxies and 
plays a crucial role as a tracer for large-scale structure (LSS) surveys of the Universe
via its 21 cm emission line of hyperfine spin-flip transition 
\citep[e.g.,][]{2004MNRAS.355.1339B,2006ApJ...653..815M,2012RPPh...75h6901P}.

However, resolving the \hi emission line individually from the distant galaxies at centimeter wavelength
needs large radio interferometers and it is also time-consuming. 
Limited by the current radio telescope resolution and sensitivity, the \hi galaxy surveys
can be only carried out within a small cosmic volume 
\citep[e.g.,][]{2001MNRAS.322..486B,2004MNRAS.350.1195M,2004MNRAS.350.1210Z,2003MNRAS.342..738L,
2005AJ....130.2598G,2007AJ....133.2569G,2007AJ....133.2087S}. 
Instead, by measuring the total \hi intensity of 
several galaxies within large voxels, known as the \hiim
\citep{2008PhRvL.100i1303C,2008PhRvL.100p1301L,
2008PhRvD..78b3529M,2008PhRvD..78j3511P,2008MNRAS.383..606W,2008MNRAS.383.1195W,
2009astro2010S.234P,2010MNRAS.407..567B,2010ApJ...721..164S,2011ApJ...741...70L,
2012A&A...540A.129A,2013MNRAS.434.1239B},
can be quickly carried out and extended to a large cosmic survey volume,
making it particularly suitable for cosmological surveys.
\citep{2015ApJ...798...40X,Zhang:2021yof,Jin:2021pcv,Wu:2021vfz,Wu:2022dgy,Wu:2022jkf,Zhang:2023gaz,
2025JCAP...01..080P}.

The \hiim technique has been investigated by analyzing the cross-correlation function 
between a Green Bank Telescope (GBT) \hiim survey and an optical galaxy survey \citep{2010Natur.466..463C}. 
Subsequent studies reported several detections of the cross-correlation power spectrum using 
\hiim data from GBT and Parkes telescopes cross-correlated with optical galaxy surveys 
\citep{2013ApJ...763L..20M,2018MNRAS.476.3382A,2017MNRAS.464.4938W,2022MNRAS.510.3495W,
2023ApJ...947...16A}.
Various ongoing \hiim experiments target the post-reionization epoch. Examples include the 
Tianlai project \citep{2012IJMPS..12..256C,2020SCPMA..6329862L,2021MNRAS.506.3455W,
2022MNRAS.517.4637P,2022RAA....22f5020S} and the Canadian Hydrogen Intensity Mapping Experiment  
\citep[CHIME,][]{2014SPIE.9145E..22B}. Upcoming \hiim experiments under development include the 
Baryonic Acoustic Oscillations from Integrated Neutral Gas Observations  
\citep[BINGO,][]{2013MNRAS.434.1239B} and the Hydrogen Intensity and Real-Time Analysis Experiment  
\citep[HIRAX,][]{2016SPIE.9906E..5XN}.
The \hiim technique is also a central focus for cosmological studies with the 
Square Kilometre Array  
\citep[SKA,][]{2015aska.confE..19S,2020PASA...37....7S,2019PhLB..79935064Z} and MeerKAT \citep{2015ApJ...803...21B,2017arXiv170906099S,2021MNRAS.501.4344L,
2021MNRAS.505.3698W,2021MNRAS.505.2039P,2023MNRAS.524.3724C}. 
Recently, a cross-correlation power spectrum detection between MeerKAT \hiim data and an 
optical galaxy survey was reported \citep{2022arXiv220601579C,2024arXiv240721626M,2024arXiv241206750C}. 
However, the \hi auto-power spectrum is still under detection with \hiim
\citep{2023arXiv230111943P,2013MNRAS.434L..46S}.

However, a significant challenge for the \hiim survey is effectively removing foreground contamination 
\citep{2015aska.confE..35W,2022MNRAS.509.2048S}. 
This issue is further complicated by various systematic effects, such as
the primary beam leakage \citep{2021MNRAS.506.5075M}, 
polarization calibration errors \citep{2016ApJ...833..289L}, 
correlated noise \citep{2018MNRAS.478.2416H,2021MNRAS.501.4344L,2021MNRAS.508.2897H}, etc.
Efforts are currently underway to address and mitigate these systematic challenges
\citep{2022ApJ...934...83N,2023MNRAS.525.5278G,2024MNRAS.527.4717I}.
Alternatively, the \hi galaxy surveys, which are free of foreground contamination, 
could be carried out with an efficiently large cosmic volume by utilizing 
next-generation radio interferometer arrays. 
Prominent examples include the SKA, 
the next-generation Very Large Array \citep[ngVLA,][]{2018ASPC..517....3M},
and the synthesis-array-upgraded Five-hundred-meter Aperture Spherical Telescope (FAST), 
referred to as the FAST Core Array \citep{2024AstTI...1...84J}.

FAST is the largest single-dish telescope in the world \citep{2013MS&E...44a2022N}, 
and since it was completed and operational in February 2020,
it has succeeded in making a number of new 
discoveries and interesting observations. As the most sensitive telescope, 
FAST shows considerable potential for cosmological studies via the \hiim technique
\citep{2017PhRvD..96f3525L,2020MNRAS.493.5854H} and is proposed with different 
drift scan cosmological surveys, 
e.g. the FAst neuTral HydrOgen intensity Mapping ExpeRiment \citep[FATHOMER,][]{2023ApJ...954..139L,2024arXiv241202582Z,2024arXiv241103988L} 
and \hi cosmology projects with the Commensal Radio Astronomy FasT Survey
\citep[CRAFTS,][]{2018IMMag..19..112L,2024arXiv241208173Y}.
Currently, the \hi galaxy surveys with FAST are still limited within 
the lower redshift \citep{2020MNRAS.493.5854H,2024SCPMA..6719511Z}.

The FAST Core Array, proposed by the FAST Operation and Development Center, 
aims to ensure that China remains competitive in the field of 
next-generation radio astronomy facilities through the 2030s. 
Another key goal is to serve as a proof-of-concept for the future FAST Array 
\citep[FASTA,][]{2023RAA....23i5005X}, 
offering crucial technical insights and identifying potential challenges that may 
arise during the implementation of FASTA.
With substantial improvements in sensitivity and angular resolution, 
the FAST Core Array offers significant potential as an ideal instrument for 
future \hi galaxy cosmological redshift surveys. 
This study employs semi-analytical simulation data to 
provide a detailed analysis of observation strategies for 
the FAST Core Array, aimed at optimizing its performance in upcoming 
\hi galaxy cosmological redshift surveys.

The rest of this paper is organized as follows. 
In \refsc{sec:simdata}, we describe the semi-analytical simulation data used in this study. 
\refsc{sec:strategy} outlines the potential observation strategies 
that could be implemented with the FAST Core Array. 
In \refsc{sec:results}, we evaluate the effects of these strategies on 
\hi galaxy surveys and present forecasts for the measurement uncertainties of the 
\hi galaxy power spectrum. 
Finally, our conclusions are summarized in \refsc{sec:conc}.

\section{Simulation data}\label{sec:simdata}

We utilize a semi-analytical simulation to estimate the galaxy number density observable with 
future FAST Core Array \hi galaxy surveys. 
This simulation is based on the JiuTian-1G (JT1G) simulation \citep{2024MNRAS.529.4958P}, 
a large-scale dark matter-only \textit{N}-body simulation developed within the framework 
of $\Lambda$ cold dark matter ($\Lambda$CDM) cosmology, designed for next-generation surveys.
The JT1G simulation was conducted using the {\sc L-Gadget-3} code 
\citep{2001NewA....6...79S,2005MNRAS.364.1105S,2012MNRAS.426.2046A}, 
within a cubic volume of $1\,h^{-1}{\rm Gpc}$ on each side. 
It achieves a particle mass resolution of $3.72 \times 10^8\,h^{-1}{\rm M_\odot}$ and 
includes 128 snapshots spanning redshifts from 128 to 0.

The galaxy catalog is constructed using the advanced semi-analytical model 
{\sc L-Galaxy} \citep{2017MNRAS.469.2626H}, modified to enhance the treatment of 
satellite galaxy disruption \citep{2024MNRAS.529.4958P}. 
For each galaxy, the {\sc L-Galaxy} model provides the total cold hydrogen gas mass. 
Following the cosmic evolution framework for the atomic and molecular phases of cold hydrogen gas, 
the \hi and molecular hydrogen (H$_2$) masses
are assigned based on the total cold gas mass and additional 
galaxy properties \citep{2009ApJ...698.1467O}.
The \hi profile is constructed via the analytical \hi and circular velocity profile models 
developed by \citet{2009ApJ...698.1467O}.
We adopt the observational frame \hi line widths, $\Delta V_{\rm o}$, 
measured at the $50$ percentile level of the peak flux densities.
Finally, the snapshots are assembled into a light cone covering a redshift range from 0 to 1, 
spanning a sky area of $20^\circ \times 20^\circ$.

The total \hi flux integrated over the galaxy velocity profile is estimated via
\citep{2017PASA...34...52M},
\begin{align}
\frac{S_{\hi}}{{\rm Jy\, km\,s}^{-1}} = \frac{(1+z)^2}{2.35 \times 10^5} 
\frac{M_{\hi}}{h^{-2}{\rm M}_{\odot}} \left(\frac{d_{\rm L}(z)}{h^{-1}{\rm Mpc}}\right)^{-2},
\end{align}
where $M_{\hi}$ is the \hi mass and $d_{{\rm L}}(z)$ is the luminosity distance to the galaxy
at redshift of $z$.

\section{Observation strategies for \hi galaxy survey}\label{sec:strategy}

The \hi galaxy detection signal-to-noise ratio (SNR) is estimated via \citep{2007AJ....133.2087S},
\begin{align}
{\rm SNR} = \frac{S_{\hi} / \Delta V_{\rm o}}{\sigma_{\rm bm}} \times 
\bigg(\frac{n_{\rm ch}}{n_{\rm bm}}\bigg)^{1/2},
\end{align}
where $S_{\hi}/ \Delta V_{\rm o}$ represent the mean \hi flux density of the galaxy, 
$n_{\rm ch}$ is the number of velocity channels the \hi profile spans, $\sigma_{\rm bm}$ is the flux density rms raised by the noise within the beam area.
Due to the large aperture of the FAST telescope, nearby galaxies may be spatially resolved. 
Additionally, with the FAST Core Array, a significant number of galaxies, particularly massive ones, 
can be resolved. 
So an extra factor $(\frac{1}{n_{\rm bm}})^{1/2}$ is applied to show the effect of galaxy resolution on detection, where $n_{\rm bm}$ is the number of beams covered by resolved galaxies, and is equal to one for unresolved galaxies. 

\begin{figure}[h]
    \centering
    \includegraphics[width=0.95\textwidth]{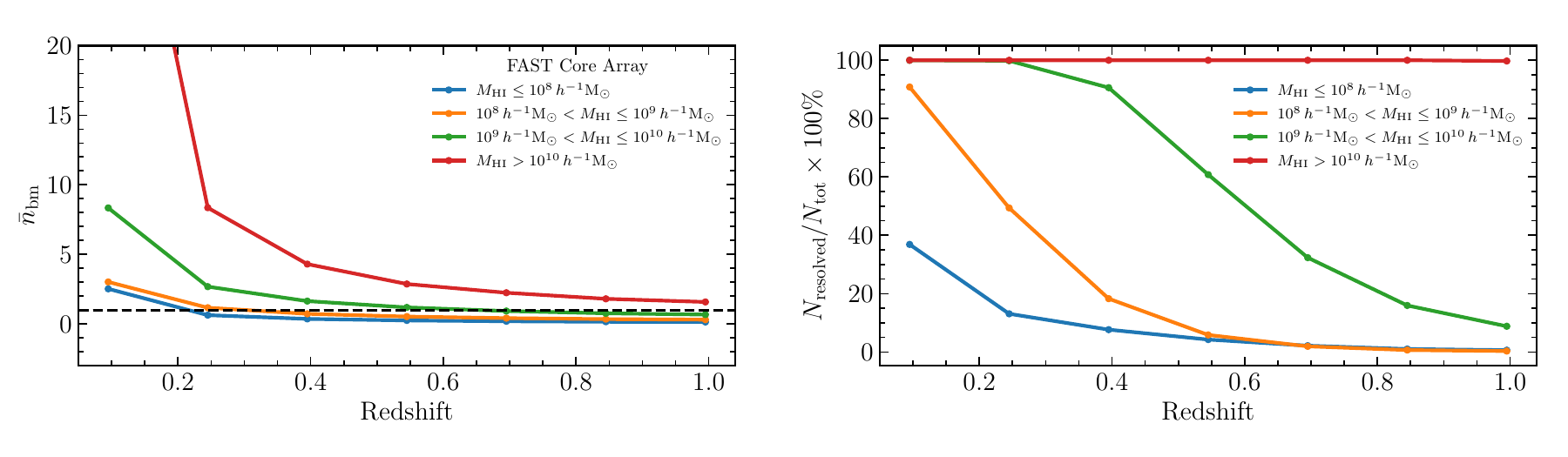}
    \caption{ The galaxy resolved by FAST Core Array as the function of redshift. 
    {\it Left panel}: the mean number of beams covered by galaxies as the function 
    of redshift. The dashed horizental line indicates $\bar{n}_{\rm bm} = 1$. 
    {\it Right panel}: the fraction of the galaxy can be resolved 
    by FAST Core Array as the function of redshift.
    In both panels, the galaxies are grouped into different \hi mass bins, and the
    corresponding results are shown in different colors, respectively. 
    }\label{fig:resolvef}
\end{figure}

The fraction of resolved galaxies may vary across different redshifts.  
Assuming a beam size of FAST Core Array, we estimate the mean number of beams covered by the galaxies 
denoted as $\bar{n}_{\rm bm}$, within different redshift bins using the simulated catalog. 
The results are presented in the left panel of  
Figure~\ref{fig:resolvef}. Additionally, we categorize galaxies based on their \hi mass, with the corresponding results shown in different colors.  
Our findings indicate that galaxies with an \hi mass greater than $10^{10}\,h^{-1}{\rm M}_\odot$  
can be well resolved up to a redshift of $z \sim 1$.  
Furthermore, we estimate the fraction of galaxies that can be resolved by the FAST Core Array,  
as shown in the right panel of Figure~\ref{fig:resolvef}.  
The results demonstrate that, except for the most massive galaxies with \hi masses exceeding  
$10^{10}\,h^{-1}{\rm M}_\odot$, the fraction of resolved galaxies decreases with increasing redshift.

The single-dish observations can be carried out with either a single FAST telescope or
$24$ individual $40\,{\rm m}$ dishes. The $\sigma_{\rm bm}$ is estimated with,
\begin{align}
\sigma^{\rm single}_{\rm bm} = \sqrt{2} \frac{k_{\rm B}T_{\rm sys}}{A_{\rm eff}} \frac{1}{\sqrt{N_{\rm dish}\Delta t \Delta \nu }},
\end{align}
where $k_{\rm B}$ is the Boltzmann constant, $T_{\rm sys}$ is the system temperature, 
$A_{\rm eff} $ is the effective aperture area, $\Delta \nu$ is the frequency resolution, 
$\Delta t$ is the total integration time within the beam area. 
For FAST single dish observation, $N_{\rm dish}=1$, and for the single dish observation with $40\,{\rm m}$
dishes, $N_{\rm dish}=24$. 

The system temperature is $T_{\text{sys}} = T_{\text{rec}} + T_{\text{sky}}$, 
where $T_{\text{rec}}$ is the receiver temperature. 
With sky area away from the Galactic plane, $T_{\rm sky}$ is modeled as \citep{2020MNRAS.493.5854H}
\begin{equation}
    T_{\text{sky}} = 2.73 + 25.2 \times (0.408/\nu_{\text{GHz}})^{2.75} K.
\end{equation}


With future FAST Core array, the \hi galaxy surveys can also be carried out with interferometer,
which consisted of either a sub-array with only the $24$ $40\,{\rm m}$ dishes, 
or the entire array including the FAST telescope. 
The flux density variance raised by noise within the synthesis beam area is estimated via, 
\begin{align}
\sigma^{\rm array}_{\rm bm} = \sqrt{2} \frac{k_{\rm B}T_{\rm sys}}{A_{\rm eff}} 
\frac{1}{\sqrt{N_{\rm dish}(N_{\rm dish} - 1)\Delta t \Delta \nu }}.
\end{align}

For the interferometry array consisted with only the $24$ $40\,{\rm m}$ dishes, 
$A_{\rm eff}$ is the effective aperture area of the $40\,{\rm m}$ dish. However,
for the interferometry array including the FAST dish, we adopted the 
mean effective aperture \citep{XuCL.in.prep.}, 
\begin{align}\label{eq:aeff}
\overline{A_{\rm eff}} = \sqrt{N'_{\text{dish}}A_{\text{eff}}^{\rm FAST}A_{\text{eff}}^{\rm 40m} + 
\frac{N'_{\text{dish}}(N'_{\text{dish}}-1)}{2} \bigg(A_{\text{eff}}^{\rm 40m}\bigg)^2},
\end{align}
where $N'_{\rm dish}= N_{\rm dish} - 1 = 24$ is the number of dishes in the $40\,{\rm m}$ array, 
$A_{\rm eff}^{\rm FAST}$ and $A_{\rm eff}^{\rm 40m}$ represent the 
effective aperture area of the FAST and $40\,{\rm m}$ dish, respectively.
Assuming equivalent aperture efficiency at 0.7, the effective collecting area is estimated to 
be approximately $50,000\, {\rm m}^2$ for FAST and 
$880\,{\rm m}^2$ for a $40\, {\rm m}$ dish. Substitution of these values into \refeq{eq:aeff} 
yields a mean effective aperture of about $35,470\,{\rm m}^2$ for the full FAST core array.

The integration time within the beam area, $\delta t$, is estimated according to different observation strategies.
In this work, we consider two different observation modes, i.e., the drift scan (DS) and on-the-fly (OTF)
observation mode, for both the FAST single dish and the FAST Core Array interferometer observation. 

\subsection{Drift scan survey}

Considering a total DS survey area of $S_{\rm area}$, the total integration time 
within the beam area is estimated via,
\begin{align}
\Delta t = \delta t \times \bigg( { N_{\rm tot} } \biggm/ { \frac{S_{\rm area}}{\delta S}} \bigg),
\end{align}
where $\delta t$ and $\delta S$ are the integration time and the survey area 
with one stripe of a drift scan observation. $S_{\rm area}/\delta S$ represent the 
number of stripes required to cover the survey area, 
and $N_{\rm tot} = t_{\rm tot} / t_{0}$ is the total number of stripes within a total observation time $t_{\rm tot}$,
where $t_0$ is the observation time spent on one drift scan stripe.
In this work, we assume that every night, we could only finish one stripe observation with $t_0 = 4\,{\rm h}$,
which spans an R.A. range of $60\,{\rm deg}$.

The sky drifts across with a speed of $\omega_{\text{e}}\text{cos}\delta$ in a drift scan survey, 
where $\omega_{\text{e}}$ $\approx$ 0.25 arcmin/s is the angular velocity of the rotation of the Earth, 
and $\delta$ is the declination of the pointing direction. 
The integration time for drifting across a pixel of the beam size is given by
\begin{equation}
    t_{\text{bm}}(z) = \frac{\theta_{\rm FWHM}(z)}{\omega_{\text{e}}\text{cos}\delta},
\end{equation}
where $\theta_{\rm FWHM}(z)$ represents the beam size at redshift $z$ 
defined as the full width at half maximum (FWHM) of the beam profile.
Considering the survey targeting at the sky close to the Zenith of the FAST site, i.e. $\delta \sim 26^\circ$, the integration time per beam can be estimated as:
\begin{equation}\label{eq:int_time}
    \frac{t_{\text{bm}(z)}}{s}=4.46\frac{\theta_{\rm FWHM}(z)}{\mathrm{arcmin}},
\end{equation}

The beam size increases with increasing redshift, resulting in a larger area covered by the scan strip and integration time per beam.
To simplify computations, the subsequent analysis assumes a constant beam size across the designated frequency band, providing a conservative prediction.

\begin{figure}
    \centering
    \includegraphics[width=0.45\textwidth]{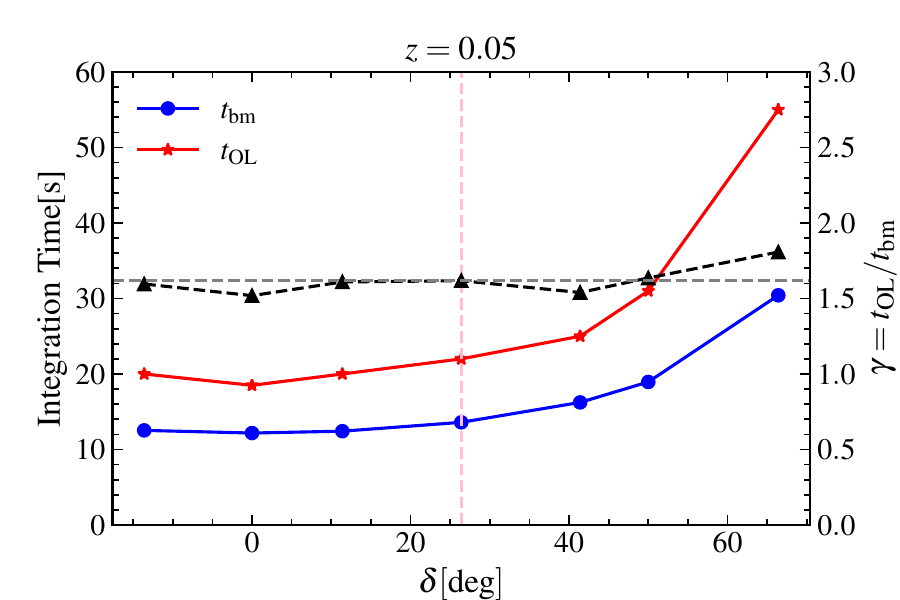}
    \includegraphics[width=0.45\textwidth]{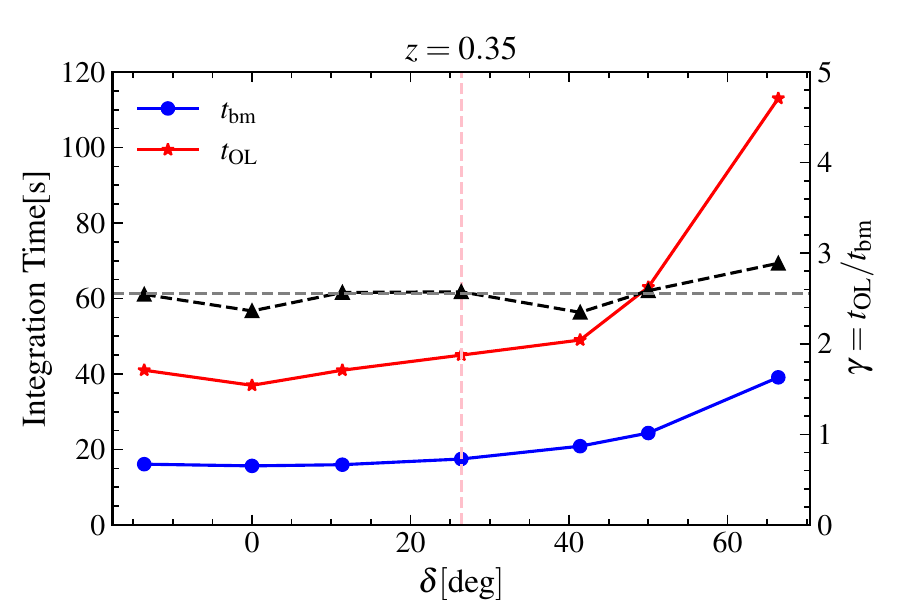}
    \caption{
    The integration time enhancement due to beam overlapping in DS observation 
    at different pointing declinations. 
    The left and right panels correspond to redshifts $z = 0.05$ and $z = 0.35$, respectively. 
    The beam-overlapping-enhanced integration time, $t_{\rm OL}$, and the initial per-beam integration time, $t_{\rm bm}$, are shown in red and blue on the left axis, while their ratio, $\gamma = t_{\rm OL}/t_{\rm bm}$, is displayed in black on the right axis. 
    The pink dashed line marks the declination corresponding to the zenith angle of the FAST site, and the gray dashed line represents the mean ratio across different pointing angles.  
    }
    \label{fig:ds_time_compare}
\end{figure}

During DS observation, rotating the FAST 19-feed array to $23.4^\circ$ ensures that the spacing between the inner beams achieves super-Nyquist sampling, meaning it is smaller than half of the full width at half maximum \citep{2018IMMag..19..112L}. This beam overlap effectively increases the integration time per sky position compared to a single-beam observation, enhancing sensitivity and improving survey efficiency.  
Using the actual beam positions and beam widths from \citep{2020RAA....20...64J}, we simulate the integration time enhancement due to beam overlapping, $t_{\rm OL}$, at redshifts $z = 0.05$ and $z = 0.35$. The results for different pointing declinations are shown in Figure~\ref{fig:ds_time_compare}.

The initial per-beam integration time, $t_{\rm bm}$, varies smoothly with the pointing angle, following an inverse proportionality to the scan speed, which scales with the cosine of the declination, peaking at the celestial equator. The beam-overlapping-enhanced integration time, $t_{\rm OL}$, exhibits a similar trend but becomes less reliable at larger pointing angles. This limitation arises because the reduced scan speed at large angles, coupled with the finite simulation resolution, prevents accurate modeling of beam overlap.  
Nevertheless, the ratio of the 19-beam to single-beam integration time can be estimated as approximately 1.6 at $z = 0.05$ and 2.6 at $z = 0.35$, demonstrating the significant integration time enhancement enabled by beam overlapping in DS mode.  

\paragraph{\it FAST L-band 19 feeds} 
For the DS survey with FAST L-band 19 feeds, we adopt the beam size 
$\theta_{\rm FWHM}(z) = 2.9 (1+z) \,{\rm arcmin}$.
The effective frequency range of the FAST L-band for \hi survey spans from $1.05\,{\rm GHz}$ to $1.40\,{\rm GHz}$,
corresponding to redshift up to $z\sim0.35$. 
The integration time per beam spans from $t_{\rm bm}(z=0.05) \simeq 13.7\,{\rm s}$ to $t_{\rm bm}(z=0.35) \simeq 17.4\,{\rm s}$.
The field of view (FoV) of the feed array can be estimated as a circular area
with a diameter of $\sim 22\, {\rm arcmin}$, since the largest angular distance of the beam center to that of the central beam is $\sim11.6\,{\rm arcmin}$ \citep{2020RAA....20...64J}. 
Therefore, with a $4$-hour drift scan observation, the 19 feeds cover a sky stripe with an area of ${\delta}S\simeq22\,{\rm arcmin}\times60\,\deg\simeq 22\,\deg^2$.
Considering the enhancement due to the beam overlapping, 
the integration time increased by $1.6$ times at $z=0.05$ and thus achive
an integration time of $\delta t\simeq1.6\times t_{\rm bm}(0.05)\simeq22\,{\rm s}$ per beam.

\paragraph{\it FAST UHF-band PAF} 
Considering the DS survey runs at the ultra-high frequency (UHF) band with the upgraded phased array feed (PAF),
we assume that the square-positioned PAF forms $10\times10$ beams covering a FoV of $36\times36\,{\rm arcmin}^2$ and works in the frequency range of $0.55$--$1.05\,{\rm GHz}$ \citep{PAF2016,2020MNRAS.493.5854H}, corresponding to the redshift range
of $0.35$--$1.58$. 
With a $4$-hour drift scan observation, the PAF covers a sky stripe with an 
area of $\delta S\simeq36\,{\rm arcmin}\times60\,\deg\simeq36\,\deg^2$ and the integration time per beam area is 
$\delta t = 10\times t_{\rm bm}(0.35) \simeq 174\,{\rm s}$,
where the factor $10$ represents the repeating observations of the $10$ beams, if they are aligned with the scanning direction.

\begin{figure}[h]
    \centering
    \includegraphics[width=0.85\textwidth]{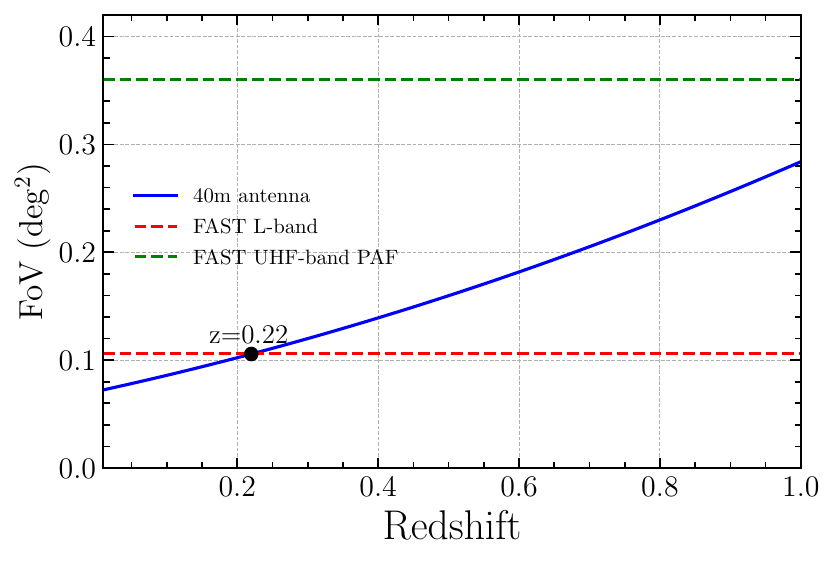}
    \caption{
    Comparison of the FoV of 40-metre antenna (blue solid), FAST L-band 19 feeds (red dashed) and FAST UHF-band PAF (green dashed).
    }
    \label{fig:FOV comparison}
\end{figure}

\paragraph{\it FAST Core Array L-band}
For the DS survey carried out with the future FAST Core Array, the FoV of the interferometer
observation is restricted by the minimal one of the FAST dish and the $40\,{\rm m}$ dish.
However, the FoV of FAST is restricted by the arrangement of 19 feeds
and the beam size variation across frequencies is negligible.
Therefore, it can simply be estimated as $\sim 0.1\, \deg^2$ using the diameter aforementioned.
While that of the 40-meter dish can be estimated with a diameter at $\sim 18\, {\rm arcmin}$ at $1.4\,{\rm GHz}$ \citep{2024AstTI...1...84J}.
As shown in \reffg{fig:FOV comparison}, the field of view (FoV) of the 40-meter dish (solid blue) is smaller than that of FAST (dashed red) up to $z = 0.22$, indicating that the 40-meter dish provides a more conservative estimate for the high-frequency sub-band, while FAST becomes the better choice for the low-frequency sub-band. Despite these differences, the survey is conducted in the L-band as a whole. More importantly, the high-frequency sub-band experiences less contamination from radio frequency interference (RFI), requiring fewer flagged data points (see \citep{2023ApJ...954..139L} for details). Consequently, we adopt the FoV of the 40-meter dish at $z = 0.05$, corresponding to a FoV diameter of $19$ arcminutes, as the reference FoV for L-band observations with the FAST Core Array interferometer.
With a $4$-hour drift scan observation, the interferometer observation at L-band
cover a sky stripe with an area of $\delta S \simeq19\,{\rm arcmin}\times60\,\deg\simeq19\,\deg^2$ and 
the integration time per beam area is $\delta t\simeq t_{\rm bm}^{40\,{\rm m}}(0.05) \simeq85\,{\rm s}$.

\paragraph{\it FAST Core Array UHF-band}

For the UHF band, the FoV is still restricted by the $40\,{\rm m}$ dish. 
We adopt the FoV of the $40\,{\rm m}$ dish at the redshift of $0.35$, 
i.e. with a diameter of $24\,{\rm arcmin}$, as the FoV for UHF-band 
interferometer DS survey. 
With a $4$-hour drift scan observation, the interferometer observation
covers a sky stripe with an area of 
$\delta S\simeq24\,{\rm arcmin}\times60\,\deg\simeq24\,\deg^2$ and 
the integration time per beam area is $\delta t \simeq t_{\rm bm}^{40\,{\rm m}}(0.35) \simeq109\,{\rm s}$.

\subsection{On-the-fly survey}
OTF observation is an efficient imaging technique that is widely used for deep-field surveys. 
By continuously collecting data as the telescope moves, OTF avoids the time-consuming stops and 
repositioning required in traditional point-by-point observations, significantly improving observational efficiency. 
In OTF mode, the telescope scans the target region along R.A. or Dec. at a constant speed \citep[e.g. $15\, \rm{arcsec}\,{\rm s}^{-1}$ along Dec. direction,][]{2020RAA....20...64J}, while the receiver records signals with high temporal resolution. Each sampling point is accurately tagged with its celestial coordinates 
based on the telescope's real-time position. 
Compared to the DS observing mode, the OTF observing mode is not limited by the 
sky’s rotation speed, allowing for more flexible scanning of localized sky regions. 
This makes it particularly suitable for deep-field surveys of specific areas.
Systematic changes that occurred from OTF observation are more benign and 
easier to correct than drifts across a single map field \citep{2007A&A...474..679M}.

The comprehensive temporal budget for an OTF observation extends beyond mere "on-source" integration, encompassing requisite "off-source" integrations and operational overhead, which includes the positional transits to and from the "OFF" reference, along with approach and transition phases \citep{2008PASJ...60..445S}. 
In an idealized scenario where overhead is negligible, the optimal "off-source" integration time scales proportionally to the "on-source" integration time, with a scaling factor of $\sqrt{N_{\rm on-off}}$, where $N_{\rm on-off}$ represents the ratio of "on-source" to "off-source" measurements \citep{2008PASJ...60..445S,2007A&A...474..679M}.
For this analysis, we posit a scenario wherein the total integration time is sufficiently ample, exemplified by an initial allocation of 100 hours, and that this entire time is effectively utilized for target area observation.

Therefore, for a target observational region $S_{\text{area}}$ and 
total survey time $t_{\rm tot}$, the integration time of the pixel by OTF mode can simply be estimated as:
\begin{equation}
    \Delta t = \frac{t_{\rm tot}}{ S_{\text{area}} / \text{FoV} }.
    \label{eq:OTF_mode}
\end{equation}

\section{results and discussion}\label{sec:results}

\subsection{Drift scan observation mode}
Due to the contamination of Global Navigation Satellite Systems (GNSS) \citep{Teunissen2017SpringerHO}, the frequency ranges of 1135-1310 MHz are mostly flagged. Therefore, the effective frequency range according to radio frequency interference (RFI) flagging for current L-band equipment is 1050-1135 MHz (low-frequency band) and 1310-1450 MHz (high-frequency band) \citep{2023ApJ...954..139L}, corresponding to the redshift range 0.25-0.35 and 0-0.08, respectively. This segmentation introduces non-uniform cosmic volume coverage, with the low-frequency sub-band probing higher redshifts (more distant galaxies) and the high-frequency sub-band targeting lower redshifts.

To effectively demonstrate the capabilities of FAST and the FAST Core Array, we predicted the observable \hi galaxy number density across varying total survey areas and integration times. 
We considered three survey area scenarios based on different zenith angles from the FAST site: Near, Optimal, and Maximum.

\begin{itemize}
    \item Near Zenith Survey: When the telescope focuses on a small area within $2^\circ$ of the FAST site's zenith, the survey area is approximately 200 deg².
    
    \item Optimal Survey: According to \cite{2020RAA....20...64J}, the system temperature of FAST reaches its minimum at a maximum zenith angle of $15^\circ$, resulting in a survey area of approximately $1600\,\rm deg^2$.
    \item  Maximum Survey: Limited by the maximum zenith angle of FAST at $40^\circ$ \citep{2018IMMag..19..112L}, the maximum survey area is approximately $4000\,\rm deg^2$.
\end{itemize}
Three total stripe counts ($N_{\rm tot} = 90$, $N_{\rm tot} = 360$, $N_{\rm tot} = 720$), approximating observation periods of three months, one year, and two years, were also considered.

\begin{figure}[t!]
    \begin{subfigure}[t]{0.48\linewidth}
    \centering
    \includegraphics[width=\textwidth]{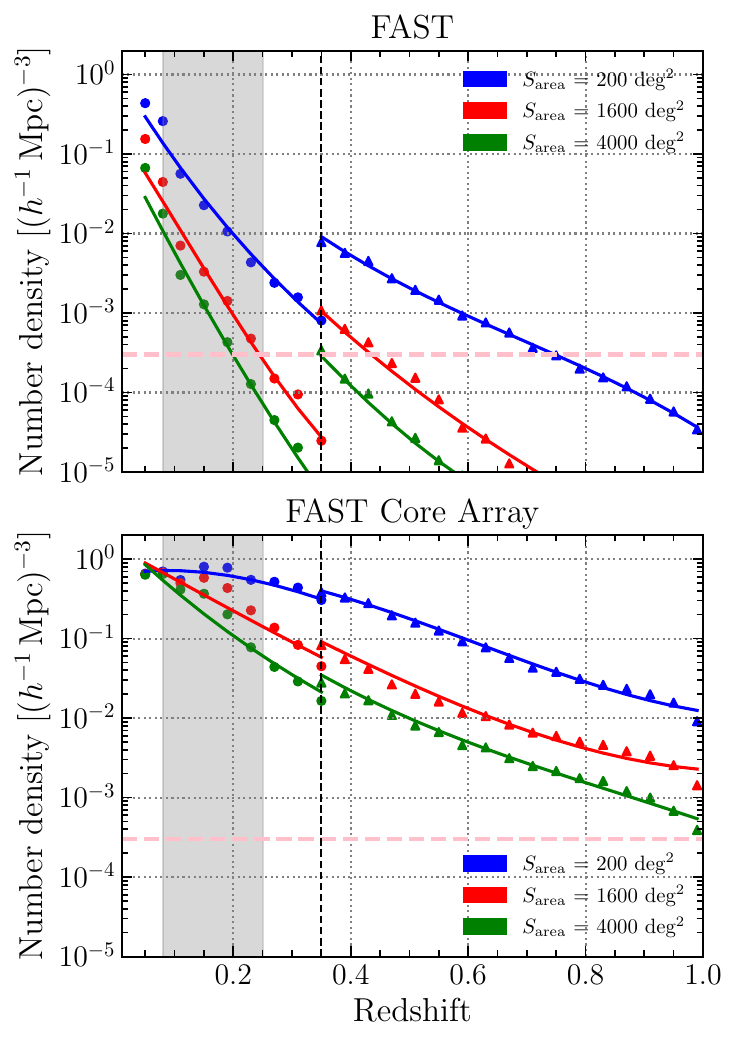}
    \caption{}\label{fig:dfscan_nd_area_no40m}
    \end{subfigure}\hfill
    \begin{subfigure}[t]{0.48\linewidth}
    \centering
    \includegraphics[width=\textwidth]{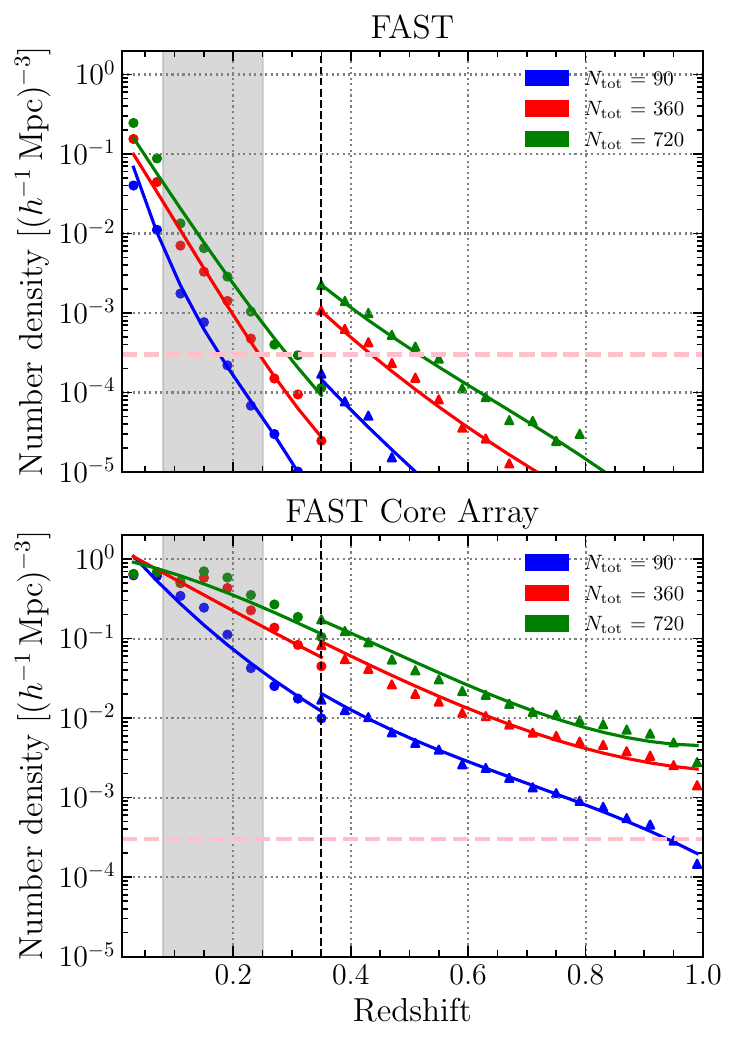}
    \caption{}\label{fig:dfscan_nd_day_no40m}
    \end{subfigure}
\caption{
Galaxy number density in the DS survey for different survey areas and survey times.  
(a) different survey areas, assuming $N_{\text{tot}} = 360$, with $S_{\rm area} = 200$, $1600$, and $4000\,\deg^2$ shown in blue, red, and green;  
(b) different survey times, assuming $S_{\text{area}} = 1600\,\deg^2$, with total observation times $N_{\text{tot}} = 90$, $360$, and $720$ shown in blue, red, and green.  
The top and bottom panels show results for the FAST single-dish and Core Array, respectively. 
Solid lines in different colors represent the corresponding best-fit curves, modeled using the expression in Equation~\ref{eq:fitting}.} The black dashed vertical line marks $z = 0.35$, separating the L-band (left) and UHF-band (right). The pink dashed horizontal line indicates a number density of $3 \times 10^{-4}\,(h^{-1}{\mpc})^{-3}$, while the gray region represents the L-band frequency gap due to GNSS RFI.

\end{figure}

The predicted \hi galaxy number densities in one-year DS survey are shown in \reffg{fig:dfscan_nd_area_no40m}.
As shown in the figure, the \hi galaxy number density detected by FAST exhibits a significant decline with increasing redshift, particularly in the L-band. 
Within the optimal survey area, the high-frequency sub-band achieves a number density of approximately $\sim 10^{-1}\,(h^{-1}\,{\rm Mpc})^{-3}$, while the low-frequency sub-band yields a substantially lower density of around $\sim 10^{-4}\,(h^{-1}\,{\rm Mpc})^{-3}$, representing a difference of approximately three orders of magnitude. However, the implementation of an upgraded PAF leads to a notable increase in the number density, reaching approximately $\sim 10^{-3}\,(h^{-1}\,{\rm Mpc})^{-3}$ at $z = 0.35$. 
Additionally, the decline in number density with increasing redshift becomes more gradual compared to the L-band. 
Indicating that FAST detectability of distant galaxies can be improved with the UHF PAF.
As we expected, the Maximum Survey obtains a lower figure than that of Optimal Survey, which is further lower than Near Zenith Survey. 
Since a smaller survey area implies a longer integration time per pixel, thereby extending the detectable galaxy distribution to higher redshifts.

For the FAST Core Array, a significant improvement is observed, particularly at higher redshifts. 
For instance, the number density in the optimal survey area increases to $\sim 10^{-1}\,(h^{-1}\,{\rm Mpc})^{-3}$at $z = 0.35$ and continues to exceed $\sim 10^{-3}\,(h^{-1}\,{\rm Mpc})^{-3}$ up to $z = 1$.
In comparison, LOWZ and CMASS sample together gives a density of order $3 \times 10^{-4}\,(h^{-1}\,{\rm Mpc})^{-3}$ at $z \sim 0.6$ \citep{2012ApJS..203...21A}

The predicted \hi galaxy number densities carried out in the Optimal Survey area are shown in \reffg{fig:dfscan_nd_day_no40m}.
As depicted in the figure, longer observation time improves the SNR of galaxy sources, resulting in significantly larger number density.
For instance, the number density of two-year observation is about an order of magnitude larger than that of three-month in both the low-frequency and high-frequency sub-bands.
FAST Core Array demonstrates substantially higher number densities than FAST, about three orders in L-band and two orders in UHF-band. 
This persistent performance across both low- and high-redshift regimes highlights the Core Array's superior suitability for galaxy surveys compared to the single-dish architecture. 

\subsection{On-the-fly observation mode}

\begin{figure}[t!]
    \begin{subfigure}[t]{0.48\linewidth}
    \centering
    \includegraphics[width=\textwidth]{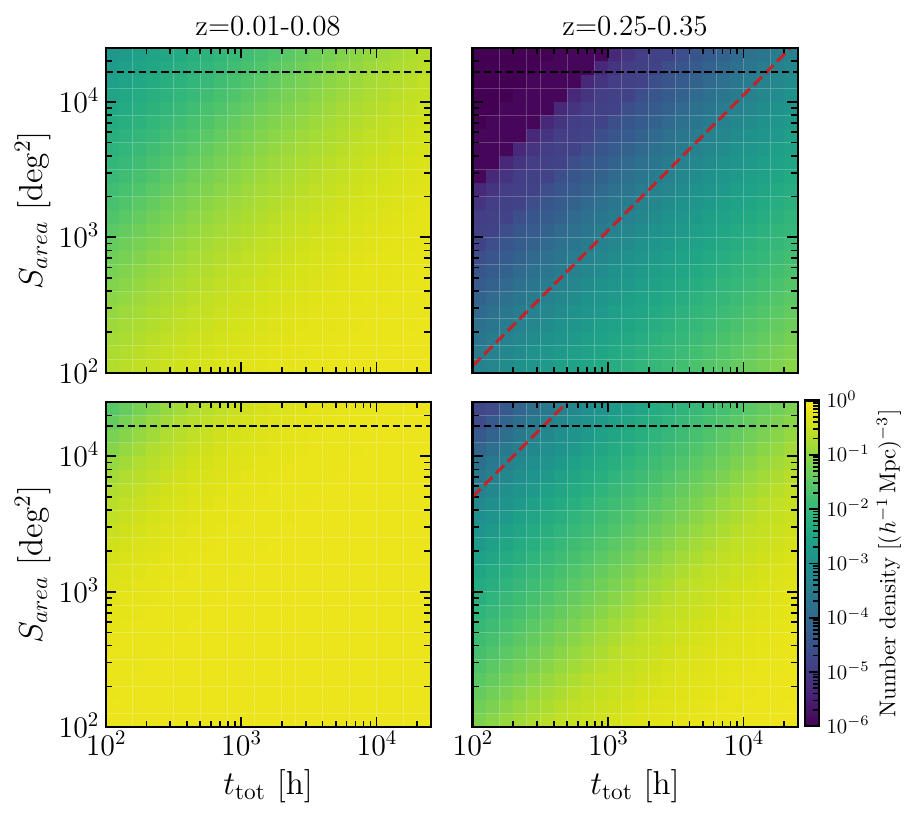}
    \caption{}\label{fig:plot_tracking_nd_L_no40m}
    \end{subfigure}\hfill
    \begin{subfigure}[t]{0.48\linewidth}
    \centering
    \includegraphics[width=\textwidth]{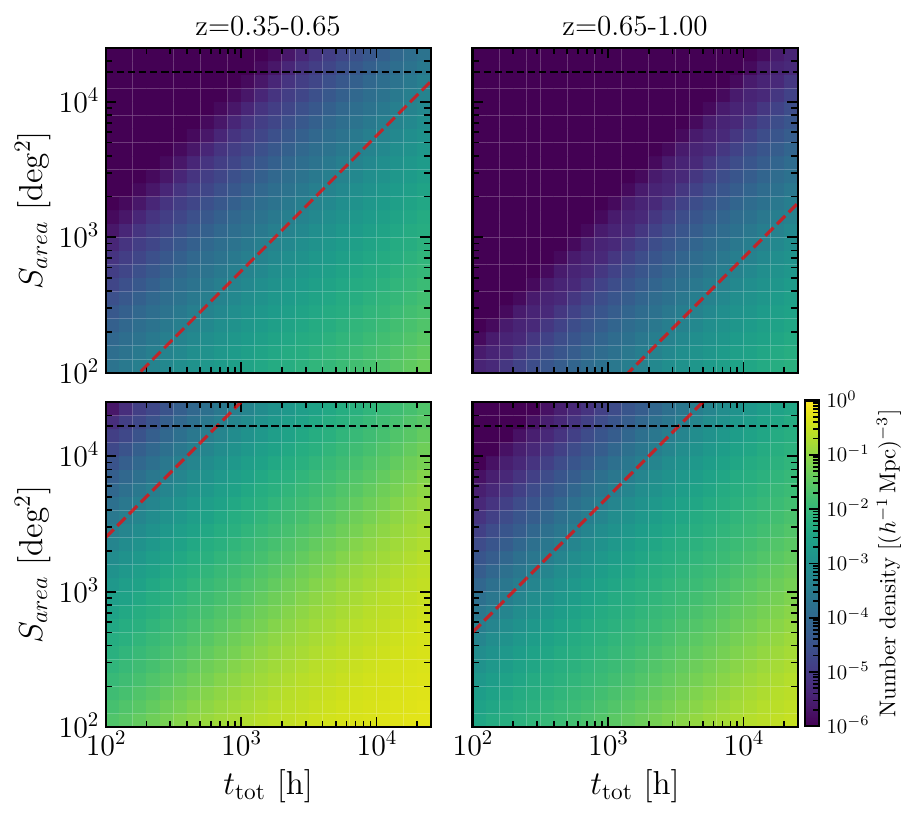}
    \caption{}\label{fig:plot_tracking_nd_UHF_no40m}
    \end{subfigure}
\caption{
   Galaxy number density in the OTF survey as a function of total survey area and integration time.
   (a) L-band results for redshift ranges $z = 0.01$–$0.08$ (left) and $z = 0.25$–$0.35$ (right).
   (b) UHF-band results for $z = 0.25$–$0.35$ (left) and $z = 0.65$–$1.00$ (right).
   The top panels show FAST single-dish results, while the bottom panels show the FAST Core Array. 
   The black dashed line marks FAST’s maximum observable survey area ($\sim 16,700\,\deg^2$) 
   after Galactic plane masking. 
   The red dashed line indicates a number density of  $3 \times 10^{-4}\,(h^{-1}\,{\rm Mpc})^{-3}$. 
} 
\end{figure}

The OTF observation mode provides a flexible scanning strategy, 
allowing surveys across various sky areas, up to FAST’s maximum observable region of 
$16\,700 \deg^2$ after masking the Galactic plane. 
\reffg{fig:plot_tracking_nd_L_no40m} presents the \hi galaxy number density 
distributions for both FAST and the Core Array in the L-band, 
covering two redshift ranges: low-$z$ ($z = 0.01$–$0.08$) and high-$z$ ($z = 0.25$–$0.35$).

At low redshift ($z = 0.01$–$0.08$), both FAST and the Core Array achieve 
sufficiently high \hi galaxy number densities, 
peaking around $10^{-1} (h^{-1}{\rm Mpc})^{-3}$ for FAST when $t_{\rm tot} = 10^3$ h 
and $S_{\rm area} = 10^3$ deg². However, the number density saturates as integration time increases,
with FAST reaching saturation at $t_{\rm tot} = 10^4$ h when surveying its full sky coverage. 
The Core Array, due to its higher sensitivity, achieves saturation with a much shorter 
integration time of $t_{\rm tot} = 10^2$ h. 
Despite the high number density, the limited cosmic volume at this redshift 
range makes it less useful for cosmological studies.

At high redshift ($z = 0.25$–$0.35$), the \hi galaxy number density decreases significantly. 
For FAST, under $t_{\rm tot} = 10^3$ h and $S_{\rm area} = 10^3$ deg², 
the density drops to $\sim 3 \times 10^{-4} (h^{-1}{\rm Mpc})^{-3}$, 
comparable to the LOWZ+CMASS sample, which serves as a reference for cosmological studies. 
The Core Array, however, maintains a much higher density of $\sim 10^{-1} (h^{-1}{\rm Mpc})^{-3}$ 
under the same conditions, making it a more effective choice for high-$z$ \hi surveys. 
This advantage stems from its superior sensitivity and larger effective aperture, 
which enhance the detection of faint \hi signals even with shorter integration times.

Figure~\ref{fig:plot_tracking_nd_UHF_no40m} extends the comparison to the UHF band, 
exploring \hi galaxy number density distributions for FAST and the Core Array at 
higher redshifts ($z = 0.35$–$0.65$ and $z = 0.65$–$1.00$) under the OTF observing mode. 
As expected, both systems exhibit lower number densities compared to their 
L-band performance due to the decreasing \hi signal strength at higher redshifts.

At intermediate redshifts ($z = 0.35$–$0.65$), the Core Array achieves a number density 
of $\sim 10^{-1} (h^{-1}{\rm Mpc})^{-3}$ with $t_{\rm tot} = 10^3$ h and $S_{\rm area} = 10^2\,\deg^2$. 
Even when scaling up to FAST’s maximum accessible survey area (black dashed line in 
Figure~\ref{fig:plot_tracking_nd_UHF_no40m}), the Core Array maintains a density 
above the LOWZ+CMASS sample’s reference value of $\sim 3 \times 10^{-4} (h^{-1}{\rm Mpc})^{-3}$. 
In contrast, under the same observing conditions, FAST’s number density falls below this 
threshold at $S_{\rm area} = 10^3\,\deg^2$, limiting its effectiveness for cosmological applications.

At higher redshifts ($z > 0.65$), the contrast between the two systems becomes even more pronounced. 
The Core Array continues to achieve detectable number densities ($> 3 \times 10^{-4} 
(h^{-1}{\rm Mpc})^{-3}$) with $t_{\rm tot} = 10^3$ h and $S_{\rm area} = 10^3\,\deg^2$, 
whereas FAST’s number density drops to $\sim 10^{-5} (h^{-1}{\rm Mpc})^{-3}$, 
falling below the critical threshold for reliable BAO detection. 
This stark disparity highlights the Core Array’s superior capability for 
high-redshift \hi surveys, driven by its advanced multi-beam 
interference suppression and aperture synthesis techniques, 
which enhance sensitivity to diffuse \hi emission.

\subsection{\hi power spectrum error}

The power spectrum is a widely used statistical tool in LSS studies. 
In this subsection, we present an error forecast for the \hi galaxy survey’s power spectrum measurement.
The primary sources of uncertainty arise from cosmic variance and shot noise. 
The fractional error in the power spectrum, evaluated over a wavenumber bin of 
width $\Delta k$, is given by \citep{1994ApJ...426...23F,2008MNRAS.383..150D},
\begin{equation}
\frac{\sigma_{P}}{P} = \sqrt{2 \frac{(2 \pi)^3}{V_{\text{eff}}} \frac{1}{4 \pi k^2 \Delta k}} \frac{P(k) + 1/n}{P(k)},
\label{eq:ps_error}
\end{equation}
where the effective survey volume is defined as
\begin{equation}
V_{\text{eff}}(k) = \int \bigg[\frac{n({r}) P(k)}{n({r}) P(k) + 1} \bigg]^{2} d^{3}{r},
\end{equation}
with $n({r})$ representing the detected galaxy number density. 
Here, $P(k)$ denotes the matter power spectrum at wavenumber $k$, 
which we compute using \camb\footnote{\url{https://camb.info}}.
In this analysis, we use $\Delta k/k = 0.125$.

To model the redshift dependence of the number density, we adopt a polynomial fitting approach:
\begin{equation}
\log_{10}\left(n(z)\right) = a z^3 + b z^2 + c z + d.
\label{eq:fitting}
\end{equation}
The function $n(z)$ was fitted to the galaxy number density for each survey strategy.
The best-fit $n(z)$ for DS surveys is shown as solid curves in Figures~\ref{fig:dfscan_nd_area_no40m} and \ref{fig:dfscan_nd_day_no40m}.
We then convert redshift into comoving distance and incorporate this relation 
into the power spectrum analysis. 
This approach ensures a more accurate estimation of power spectrum uncertainties 
across different redshift bins.

\subsubsection{Drift scan observation}

\begin{figure}[t]
    \begin{subfigure}[t]{0.46\textwidth}
    \centering
    \includegraphics[width=\textwidth]{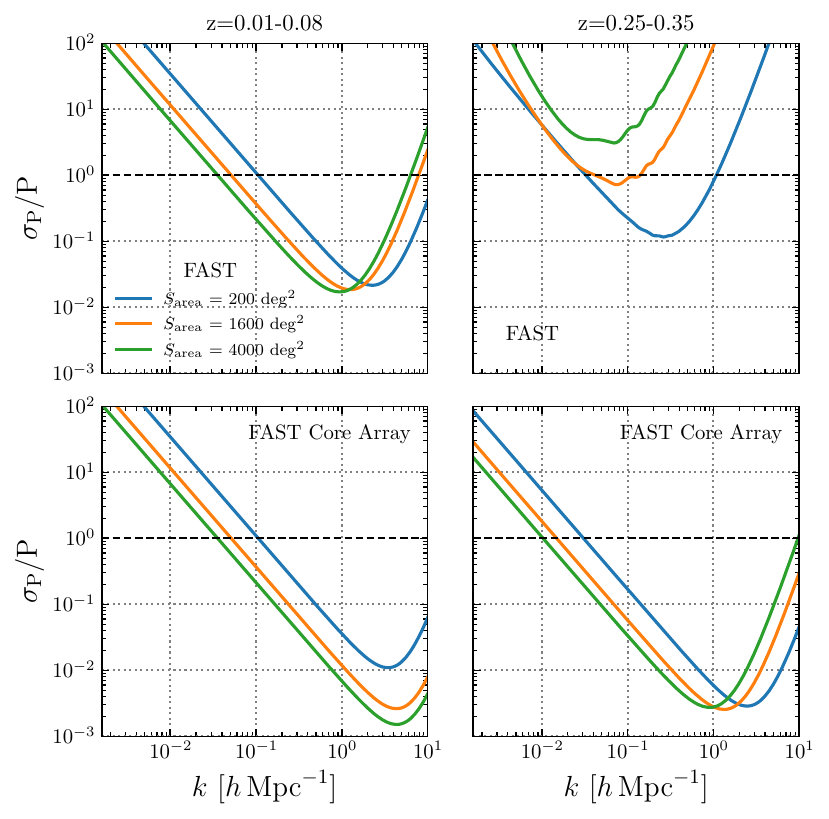}
    \caption{}\label{fig:plot_pserror_ds_nd_area_L}
    \end{subfigure}\hfill
    \begin{subfigure}[t]{0.46\textwidth}
    \centering
    \includegraphics[width=\textwidth]{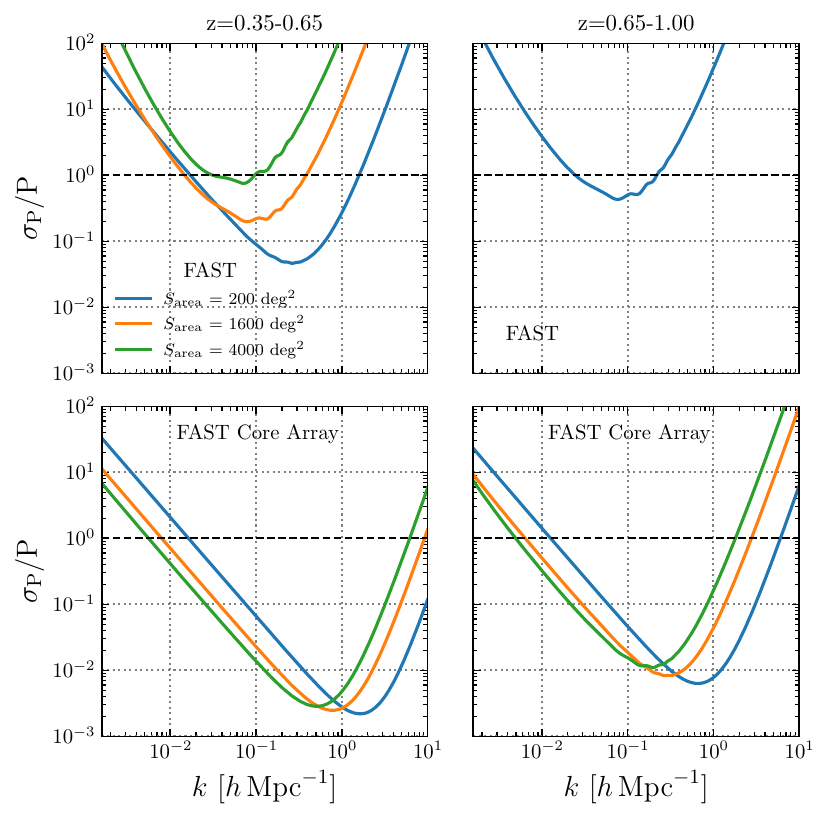}
    \caption{}\label{fig:plot_pserror_ds_nd_area_UHF}
    \end{subfigure}
    \caption{
    Fractional error in the power spectrum for the DS observation
    with $N_{\text{tot}} = 360$ for survey areas $S_{\text{area}} = 200$, 1600, and 4000 $\text{deg}^2$.
    (a) L-band results with the left and right panels correspond to redshift ranges 
    $z = 0.01$--$0.08$ and $z = 0.25$--$0.35$, separated by GNSS RFI contamination. 
    (b) UHF-band results. The left and right panels correspond to redshift ranges 
    $z = 0.35$--$0.65$ and $z = 0.65$--$1.00$.  
    The top and bottom panels show results for the FAST single-dish and Core Array, respectively. 
    The black dashed horizontal line marks $\sigma_{P}/P = 1$.
    }
\end{figure}

The fractional errors of the power spectrum for the DS survey are shown in Figure~\ref{fig:plot_pserror_ds_nd_area_L} (L-band) and Figure~\ref{fig:plot_pserror_ds_nd_area_UHF} (UHF-band). We fix the total observation time at $N_{\rm tot} = 360$ and present results for survey areas $S_{\rm area} = 200$, $1600$, and $4000\, \deg^2$, shown in blue, orange, and green, respectively. The top and bottom panels in each figure correspond to results from the FAST single-dish survey and the FAST Core Array, respectively.  

At low redshift ($0.01 < z < 0.08$), both the FAST single-dish and Core Array surveys achieve high galaxy number densities, making cosmic variance the dominant source of uncertainty in the power spectrum measurement. However, due to the limited cosmic volume at these redshifts, the power spectrum measurement is primarily restricted to small scales, with sensitivity concentrated at $k \gtrapprox 0.1\,h\,{\rm Mpc}^{-1}$.  

As redshift increases, the fractional errors in the power spectrum for the FAST single-dish survey grow significantly, particularly in the higher-redshift range of the L-band ($0.25 < z < 0.35$) and across the UHF-band ($0.35 < z < 1.00$). This increase is primarily due to a sharp decline in the number density of detected \hi galaxies, which reduces the statistical power of the measurement. For $S_{\rm area} = 200\,\deg^2$, the errors exceed unity ($\sigma_P/P > 1$) at large scales in the UHF-band, making power spectrum constraints challenging.  

In contrast, the FAST Core Array provides significantly improved power spectrum measurements at higher redshifts. With its higher sensitivity and angular resolution, the Core Array achieves a sufficient \hi galaxy number density even at large distances. At high redshifts ($z > 0.25$), the increased survey volume further reduces cosmic variance, enabling robust power spectrum detection across a wider range of scales. Notably, the Core Array is capable of probing large scales ($k \sim 0.1\,h\,{\rm Mpc}^{-1}$) while maintaining sensitivity at small scales ($k > 1\,h\,{\rm Mpc}^{-1}$). 


In addition, we investigate the fractional errors in the \hi power spectrum for different DS survey observation times. 
The results are presented in Figure~\ref{fig:plot_pserror_ds_nd_day_L} for the L-band and Figure~\ref{fig:plot_pserror_ds_nd_day_UHF} for the UHF-band, both assuming a fixed survey area of $S_{\rm area} = 1600\,\deg^2$. 

The top panels of Figures~\ref{fig:plot_pserror_ds_nd_day_L} and \ref{fig:plot_pserror_ds_nd_day_UHF} show the results for the FAST single-dish observation. 
As the total observation time increases up to $N_{\rm tot} = 720$, equivalent to two years of continuous observation, the single-dish survey achieves limited power spectrum detection up to redshift $z \sim 0.65$. 
However, the measurements remain largely dominated by cosmic variance and shot noise, significantly limiting their statistical precision, particularly at high redshift. 
At $z > 0.35$, the signal-to-noise ratio deteriorates, making it challenging to extract reliable cosmological information from the observed \hi fluctuations. 

In contrast, the bottom panels of Figures~\ref{fig:plot_pserror_ds_nd_day_L} and \ref{fig:plot_pserror_ds_nd_day_UHF} illustrate the results for the FAST Core Array. 
Due to its superior sensitivity and angular resolution, the Core Array enables robust power spectrum measurements extending up to $z \sim 1$. 
Even at higher redshifts, where the \hi signal is weaker and the number density decreases, the Core Array maintains a significantly lower fractional error compared to the single-dish mode. 
This enhanced capability makes it a powerful tool for probing large-scale structure evolution and constraining cosmological parameters across a wide redshift range.

\subsubsection{On-the-fly observation}

\begin{figure}[t]
    \begin{subfigure}[t]{0.46\textwidth}
    \centering
    \includegraphics[width=\textwidth]{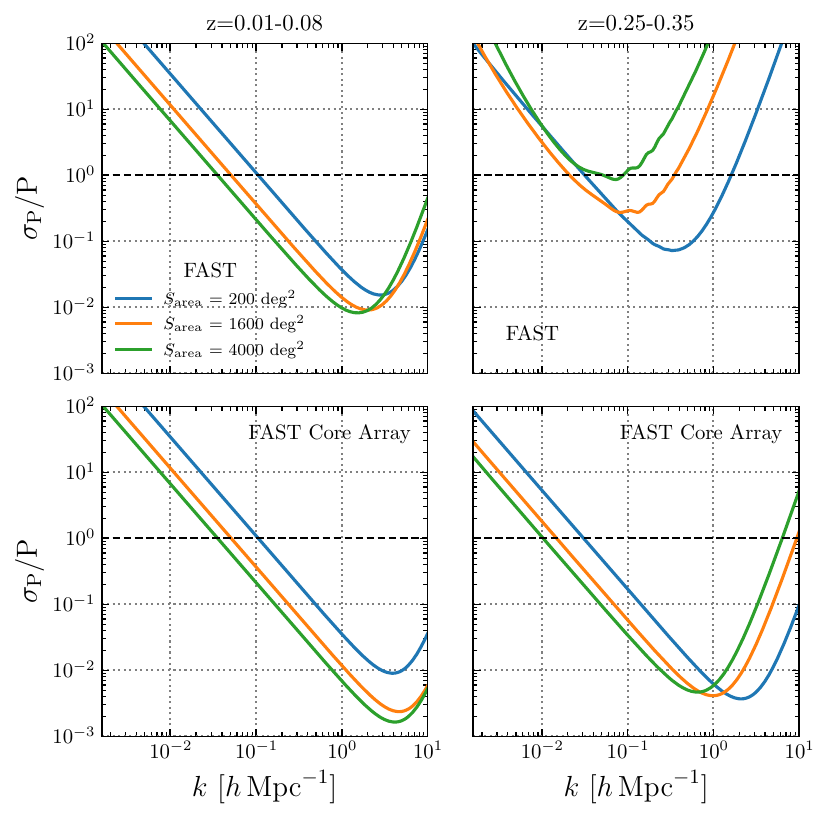}
    \caption{}\label{fig:plot_pserror_otf_nd_area_L}
    \end{subfigure}\hfill
    \begin{subfigure}[t]{0.46\textwidth}
    \centering
    \includegraphics[width=\textwidth]{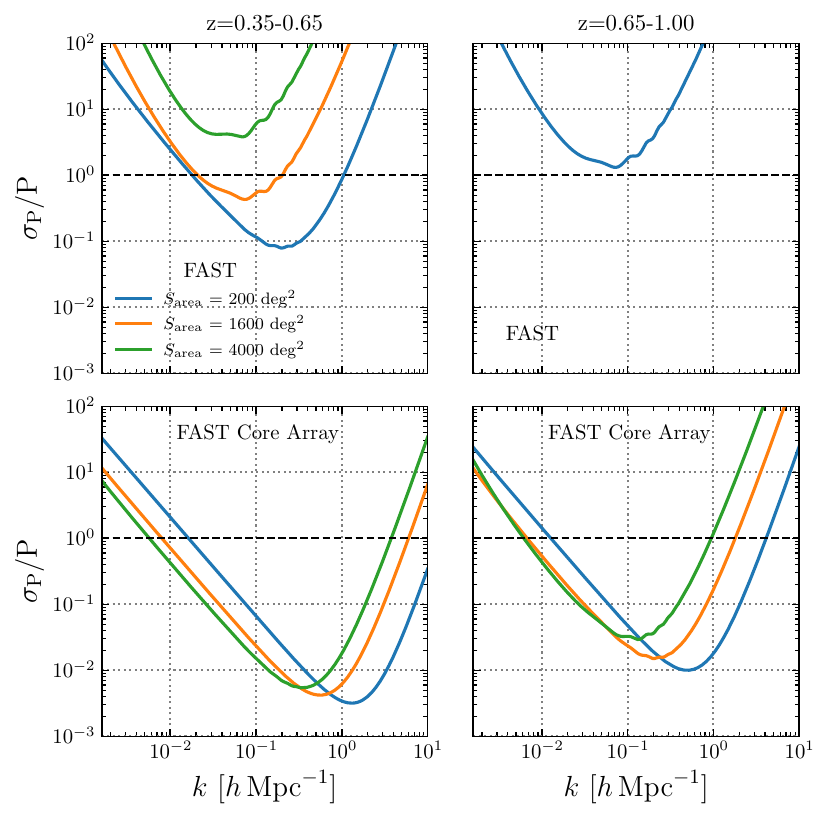}
    \caption{}\label{fig:plot_pserror_otf_nd_area_UHF}
    \end{subfigure}
    \caption{
    Fractional error in the power spectrum for the OTF survey with $t_{\rm tot} = 1000\,{\rm h}$ for survey areas of $S_{\rm area}$ = 200, 1600, and 4000 $\deg^2$. 
    (a) L-band results with the left and right panels correspond to redshift ranges 
    $z = 0.01$--$0.08$ and $z = 0.25$--$0.35$, separated by GNSS RFI contamination. 
    (b) UHF-band results. The left and right panels correspond to redshift ranges 
    $z = 0.35$--$0.65$ and $z = 0.65$--$1.00$.  
    The top and bottom panels show results for the FAST single-dish and Core Array, respectively. 
    The black dashed horizontal line marks $\sigma_{P}/P = 1$.  
    }
\end{figure}

\begin{figure}[t]
    \begin{subfigure}[t]{0.46\textwidth}
    \centering
    \includegraphics[width=\textwidth]{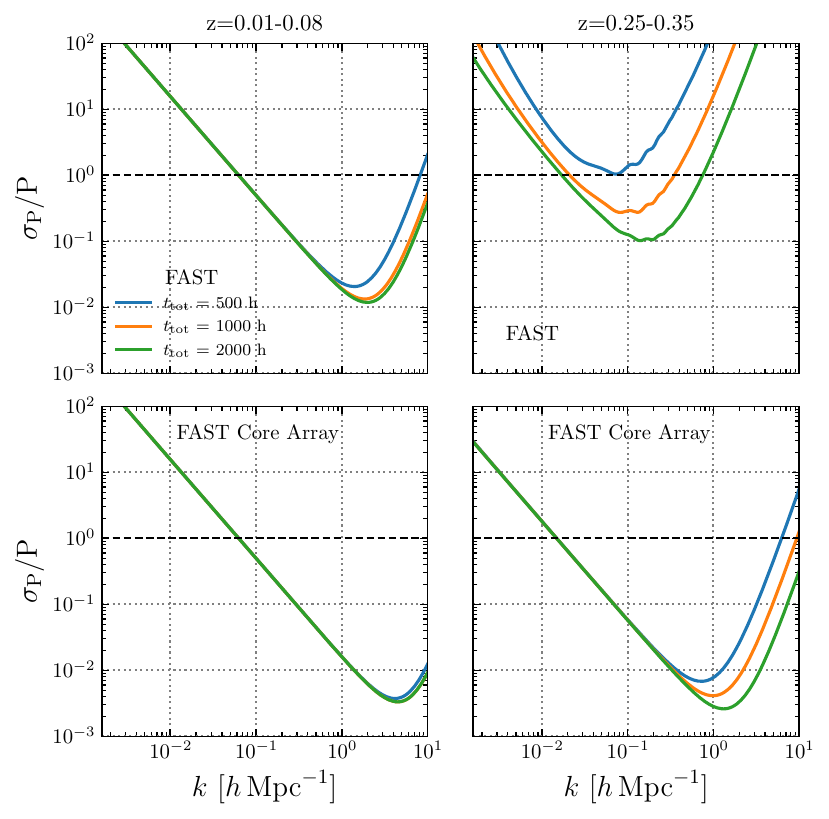}
    \caption{}\label{fig:plot_pserror_otf_nd_h_L}
    \end{subfigure}\hfill
    \begin{subfigure}[t]{0.46\textwidth}
    \centering
    \includegraphics[width=\textwidth]{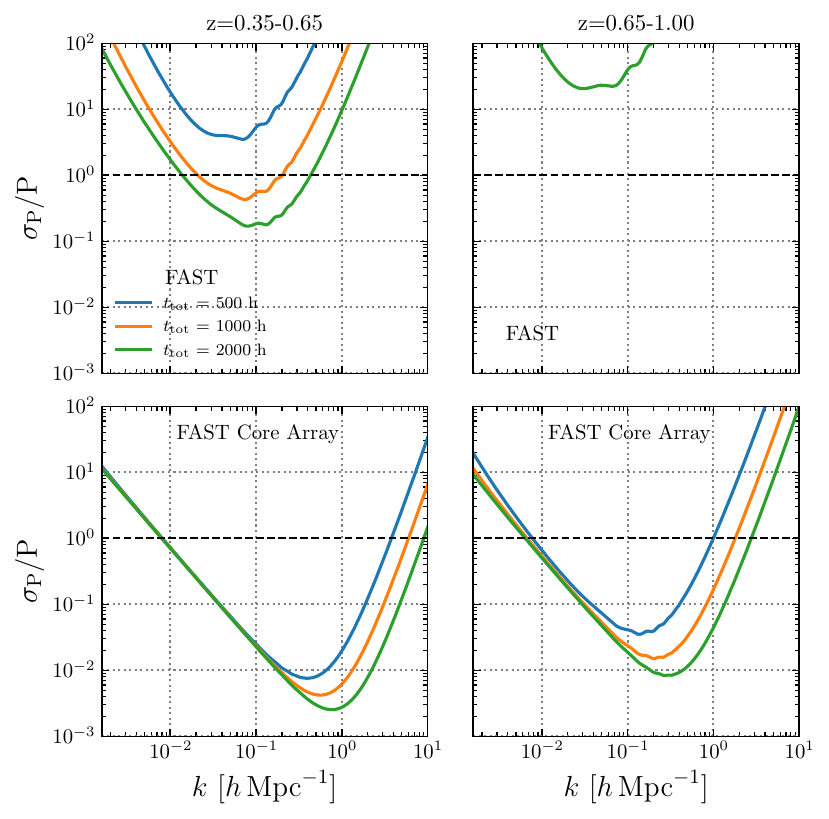}
    \caption{}\label{fig:plot_pserror_otf_nd_h_UHF}
    \end{subfigure}
    \caption{
    Fractional error in the power spectrum for the OTF survey with $S_{\rm area} = 1600\,{\deg}^2$ for total observation times of $t_{\rm tot} = 500$, 1000, and 2000 ${\rm h}$. 
    (a) L-band results with the left and right panels correspond to redshift ranges 
    $z = 0.01$--$0.08$ and $z = 0.25$--$0.35$, separated by GNSS RFI contamination. 
    (b) UHF-band results. The left and right panels correspond to redshift ranges 
    $z = 0.35$--$0.65$ and $z = 0.65$--$1.00$.  
    The top and bottom panels show results for the FAST single-dish and Core Array, respectively. 
    The black dashed horizontal line marks $\sigma_{P}/P = 1$.  
    }
\end{figure}

We further investigate the fractional errors of the \hi power spectrum using OTF observations, 
considering different survey areas and observation times. 

We first fix the total observation time at $t_{\rm tot} = 1000$ h and vary the survey area 
among $S_{\rm area} = 200$, $1600$, and $4000\,\deg^2$. The corresponding results for the 
L-band and UHF-band are presented in Figures~\ref{fig:plot_pserror_otf_nd_area_L} and 
\ref{fig:plot_pserror_otf_nd_area_UHF}, respectively.

Similar to the DS observations, in the redshift range $0.01 \leq z \leq 0.08$, both the FAST 
single-dish and Core Array configurations yield significant detections of the \hi power spectrum. 
Due to the limited cosmic volume at these low redshifts, most detections occur at small scales 
($k \gtrsim 0.1\,h\,{\rm Mpc}^{-1}$), where the sample variance is minimized.

At higher redshifts, the fractional errors in the \hi power spectrum increase substantially for the FAST 
single-dish observation. The decrease in detected \hi galaxy number density results in shot-noise-dominated 
uncertainties, making power spectrum measurements highly sensitive to galaxy shot noise rather than 
cosmological signals. Consequently, expanding the survey area does not significantly improve the measurement 
precision for the FAST single-dish mode.

In contrast, the FAST Core Array achieves a notable improvement in power spectrum measurements, particularly at 
higher redshifts. Its enhanced sensitivity and spatial resolution allow for more reliable detections of the \hi 
power spectrum, highlighting its advantage in probing large-scale structure evolution beyond $z > 0.35$.

Next, we examine the effect of varying the total observation time on power spectrum measurements. 
We fix the survey area at $S_{\rm area} = 1600\,\deg^2$ and consider total observation times of 
$t_{\rm tot} = 500$, $1000$, and $2000$ h. The results for the L-band and UHF-band are displayed 
in Figures~\ref{fig:plot_pserror_otf_nd_h_L} and \ref{fig:plot_pserror_otf_nd_h_UHF}, respectively.

At low redshifts ($0.01 \leq z \leq 0.08$), increasing the integration time provides little improvement 
in power spectrum measurements. This is because the \hi galaxy number density saturates at these redshifts, 
and the dominant source of uncertainty is cosmic variance rather than shot noise.

At higher redshifts ($z > 0.35$), increasing the observation time leads to a noticeable reduction in 
fractional errors, especially for the FAST single-dish mode. However, despite this improvement, the 
single-dish configuration still struggles to provide sufficiently precise power spectrum measurements 
for strong cosmological constraints. For the FAST Core Array, increasing the integration time significantly enhances detection precision. 

Our analysis demonstrates that survey strategy plays a crucial role in optimizing \hi power spectrum 
measurements. At low redshifts, both the FAST single-dish and Core Array configurations perform well, 
with small-scale detections benefiting from high galaxy number density. However, at higher redshifts, 
the FAST Core Array significantly outperforms the single-dish mode, particularly when increasing the 
survey area and observation time. While longer integration times enhance sensitivity, their impact 
diminishes beyond a certain threshold, emphasizing the need for an optimal balance between survey 
area, integration time, and detection strategy in future \hi galaxy surveys.

\section{Summary and Conclusion}\label{sec:conc}

In this work, we have explored various strategies for conducting \hi galaxy redshift surveys with the FAST telescope and its proposed Core Array configuration. Through semi-analytical simulations, we assessed the impact of different observational modes, survey areas, and integration times on the detectability of \hi galaxies and the accuracy of power spectrum measurements.  

We assume that surveys using the L-band and UHF-band are conducted separately.
Since the beam size varies across each frequency band, we adopt the beam size at the highest frequency of each band for analysis,
corresponding to a redshift of 0.05 for the L-band and 0.35 for the UHF-band.
As redshift increases, the beam size grows, leading to a larger scan strip coverage and increased integration time per beam.
To simplify computations, we assume a constant beam size across each frequency band in the subsequent analysis,
providing a conservative prediction.

Comparing the FAST single-dish mode with the FAST Core Array, we find that the single-dish mode achieves significant \hi detections at low redshifts ($z \lesssim 0.35$), but its performance deteriorates at higher redshifts due to its lower sensitivity and larger beam size, which result in increased shot noise contamination. In contrast, the FAST Core Array, with its improved sensitivity and angular resolution, maintains a higher \hi galaxy number density, particularly at $z > 0.35$, making it better suited for cosmological studies. The Core Array remains effective for BAO measurements at high redshifts, while the single-dish mode falls below the critical detection threshold.  

Regarding observational strategies, we compare the drift scan (DS) and on-the-fly (OTF) modes. The DS mode is constrained by Earth's rotation, which limits flexibility in sky coverage but enables efficient large-area surveys with minimal operational overhead. In contrast, the OTF mode provides greater flexibility, particularly for targeted deep-field surveys, allowing more uniform integration times and optimized sky coverage. Although OTF introduces additional observational overhead, its efficiency in deep-field mapping makes it a valuable approach for high-precision studies.  

The performance of both FAST configurations varies significantly with redshift. At low redshifts ($z \sim 0.01$–$0.08$), both the single-dish and Core Array configurations achieve high \hi number densities, making cosmic variance the dominant source of uncertainty in power spectrum measurements. Due to the limited cosmic volume at these redshifts, detections are primarily constrained to small scales ($k \gtrsim 0.1\,h\,{\rm Mpc}^{-1}$). At higher redshifts ($z > 0.35$), the FAST Core Array significantly outperforms the single-dish mode. The Core Array maintains detectable \hi number densities and provides robust power spectrum constraints, whereas the single-dish observations become dominated by shot noise. Increasing the survey area does little to mitigate this limitation for the single-dish mode, whereas it improves the performance of the Core Array.  

Our results highlight the importance of survey strategy in optimizing \hi galaxy redshift surveys. The FAST single-dish mode is well-suited for low-redshift \hi studies, particularly in wide-area surveys. However, for precision cosmology at higher redshifts, the FAST Core Array provides a substantial improvement, offering better sensitivity and enabling large-scale structure studies up to $z \sim 1$. The choice between drift scan and OTF mode depends on the specific scientific objectives. Drift scan surveys are ideal for wide-area coverage, while OTF observations are better suited for targeted deep-field studies.  

Future work should focus on refining survey designs based on real observational constraints, including potential instrumental systematics and calibration strategies. Additionally, integrating \hi intensity mapping techniques with resolved \hi galaxy detections could further enhance the scientific output of FAST and its Core Array, providing valuable constraints on the evolution of large-scale structure and dark energy.

\begin{acknowledgements}
We sincerely thank Yun Liu and Wenxiang Pei for their valuable discussions
and support in providing simulation data.
We acknowledge the support of the National SKA Program of China 
(Nos.~2022SKA0110200, 2022SKA0110203),
the National Natural Science Foundation of China (Nos. 12473091,12473001,), 
111 Project (No. B16009), and the NSFC International (Regional) Cooperation and Exchange Project (No. 12361141814).
We express our gratitude to ChatGPT for its assistance in refining the manuscript and enhancing the code to improve execution efficiency.
\end{acknowledgements}

\bibliography{references}{}
\bibliographystyle{aasjournal}

\label{lastpage}

\end{document}